\documentclass{article}
\usepackage{spconf,amsmath,graphicx,multirow,booktabs,amssymb,gensymb}

\title{Spatial and spectral deep attention fusion for multi-channel speech separation using deep embedding features}
\name{Cunhang Fan$^{1,3}$, Bin Liu$^1$, Jianhua Tao$^{1,2,3}$, Jiangyan Yi$^{1}$, Zhengqi Wen$^1$}
\address{ $^1$NLPR, Institute of Automation, Chinese Academy of Sciences, Beijing, China\\
	$^2$CAS Center for Excellence in Brain Science and Intelligence Technology, Beijing, China\\
	$^3$School of Artificial Intelligence, University of Chinese Academy of Sciences, Beijing, China}
\begin{document}
%
\maketitle
\begin{abstract}
Multi-channel deep clustering (MDC) has acquired a good performance for speech separation. However, MDC only applies the spatial features as the additional information. So it is difficult to learn mutual relationship between spatial and spectral features. Besides, the training objective of MDC is defined at embedding vectors, rather than real separated sources, which may damage the separation performance. In this work, we propose a deep attention fusion method to dynamically control the weights of the spectral and spatial features and combine them deeply. In addition, to solve the training objective problem of MDC, the real separated sources are used as the training objectives. Specifically, we apply the deep clustering network to extract deep embedding features. Instead of using the unsupervised K-means clustering to estimate binary masks, another supervised network is utilized to learn soft masks from these deep embedding features. Our experiments are conducted on a spatialized reverberant version of WSJ0-2mix dataset. Experimental results show that the proposed method outperforms MDC baseline and even better than the oracle ideal binary mask (IBM).
\end{abstract}

\begin{keywords}
Multi-channel deep clustering, speech separation, deep attention fusion, deep embedding features
\end{keywords}

\vspace{-2ex}
\section{Introduction}
\vspace{-1ex}
\label{sec:intro}

Speech separation is known as the cocktail party problem \cite{O2015Attentional}, which aims to estimate the target sources from a noisy mixture. To address this problem, there are many works have been done and made significant advances, such as deep clustering (DC) \cite{Hershey2016Deep,isik2016single}, permutation invariant training (PIT) \cite{Yu2017Permutation,Kolbaek2017Multitalker} and Conv-TasNet \cite{luo2019conv}. They all do not use the spatial information because they are monaural speech separation methods. As for the multiple microphones, they contain the directional information of each source. Therefore, the spatial features can be leveraged to the multi-channel speech separation. 


Recently, to utilize the spatial information, many works have been done for multi-channel speech separation \cite{wang2018integrating,wang2018combining,togami2019spatial,gu2019neural}. Multi-channel deep clustering (MDC) \cite{wang2018multi} extends the DC to multi-channel. DC \cite{Hershey2016Deep} is a single channel speech separation technique. It trains a bidirectional long-short term memory (BLSTM) network to map the mixed spectrogram into an embedding space. At testing stage, the embedding vector of each time-frequency (T-F) bin is clustered by K-means to obtain binary masks. Different from DC, MDC uses the interchannel phase differences (IPDs) \cite{chen2018multi} as the additional spatial features to the separation model. In other words, MDC applies not only spectral but also the spatial features as the input for better separation. Although MDC can separate the mixture well, there are still two limitations. Firstly, MDC only uses the spatial features as the additional information, which is difficult to learn mutual relationship between spatial and spectral features. Secondly, the training objective of MDC is defined at the embedding vectors, rather than real separated sources. These embedding vectors do not necessarily imply the perfect separation of sources in signal space.
In this paper, we propose a spatial and spectral deep attention fusion method for multi-channel speech separation using deep embedding features. Different from MDC only using the IPDs as the additional features, we propose a deep attention fusion algorithm to combine the spectral and spatial features deeply. Therefore, the separation model can dynamically control the weights of the spectral and spatial features. In addition, to address the training objective problem of MDC, motivated by our previous work \cite{fan2019discriminative}, we apply the deep embedding features for multi-channel speech separation. Specifically, the MDC network is utilized to extract deep embedding features. Instead of using the unsupervised K-means clustering algorithm to estimate binary masks, the supervised utterance-level PIT (uPIT) \cite{Kolbaek2017Multitalker} network is applied to learn soft masks from these deep embedding features. Therefore, the separation model can use the real separated sources as the training objective. Finally, to reduce the distance between the same speakers and increase the distance between different speakers, the discriminative learning \cite{fan2018Utterance} is utilized to fine-tune the separation model.


To summarize, the main contribution of this paper is two-fold. Firstly, we propose a deep attention fusion algorithm to combine the spectral and spatial features deeply. Secondly, the MDC is applied to extract deep embedding features. And another supervised uPIT network is used to learn target masks instead of the unsupervised K-means clustering. Therefore, the separation model can use the real separated sources as the training objective.


The rest of this paper is organized as follows. Section 2 presents the multi-channel deep clustering. The  proposed method is stated in section 3. Section 4 shows detailed experiments and results. Section 5 draws conclusions.

\vspace{-2ex}
\section{Multi-channel deep clustering}
\vspace{-1ex}

The aim of single-channel deep clustering (DC) \cite{Hershey2016Deep,isik2016single} is to map the mixture spectrogram into a high-dimensional embedding \(V\) for each T-F bin by a deep neural network (DNN). The loss function of DC is defined as follows:
\begin{equation}
\begin{split}
J_{DC} &=||VV^T-BB^T||_F^2\\
&=||VV^T||_F^2-2||V^TB||_F^2+||BB^T||_F^2\\
\end{split}
\label{eq1}
\end{equation}
where \(B\in{\mathbb{R}^{TF\times{S}}}\) is the source membership function for each T-F bin, i.e.,\(B_{tf,s}=1\), if source \(s\) has the highest energy at time \(t\) and frequency \(f\) compared to the other sources. Otherwise, \(B_{tf,s}=0\). \(S\) is the number of sources. \(||*||_F^2\) is the squared Frobenius norm.

The difference between single-channel DC and multi-channel DC (MDC) is the input features. As for the single-channel DC, only the mixture spectrogram \(|Y(t,f)|\) is used as the input feature: \(\zeta_{DNN}=\{|Y(t,f)|\}\). As for the MDC, the phase difference between two microphones, \(\theta_i(t,f)\) (\(i\) is the index of a microphone pair), is applied as an additional input feature as follows:
\begin{equation}
\begin{split}
\zeta_{DNN}=\{|Y(t,f)|; cos\theta_i(t,f); sin\theta_i(t,f)\}
\end{split}
\label{eq2}
\end{equation}
Besides, when the number of microphone \(N_m > 2\), MDC firstly chooses a reference microphone and each pair \(\theta_i(t,f)\) is computed between a reference and non-reference microphone. Therefore, there will be \(N_m-1\) embeddings. When these embeddings are stacked at each T-F bin, the K-means clustering is applied to estimate binary masks. Finally, these masks are utilized to the reference microphone signal for separation.

\vspace{-2ex}
\section{The proposed separation method}
\vspace{-1ex}

\begin{figure}[t]
	\centering
	\begin{minipage}[t]{0.32\textwidth}
		\includegraphics[width=\linewidth]{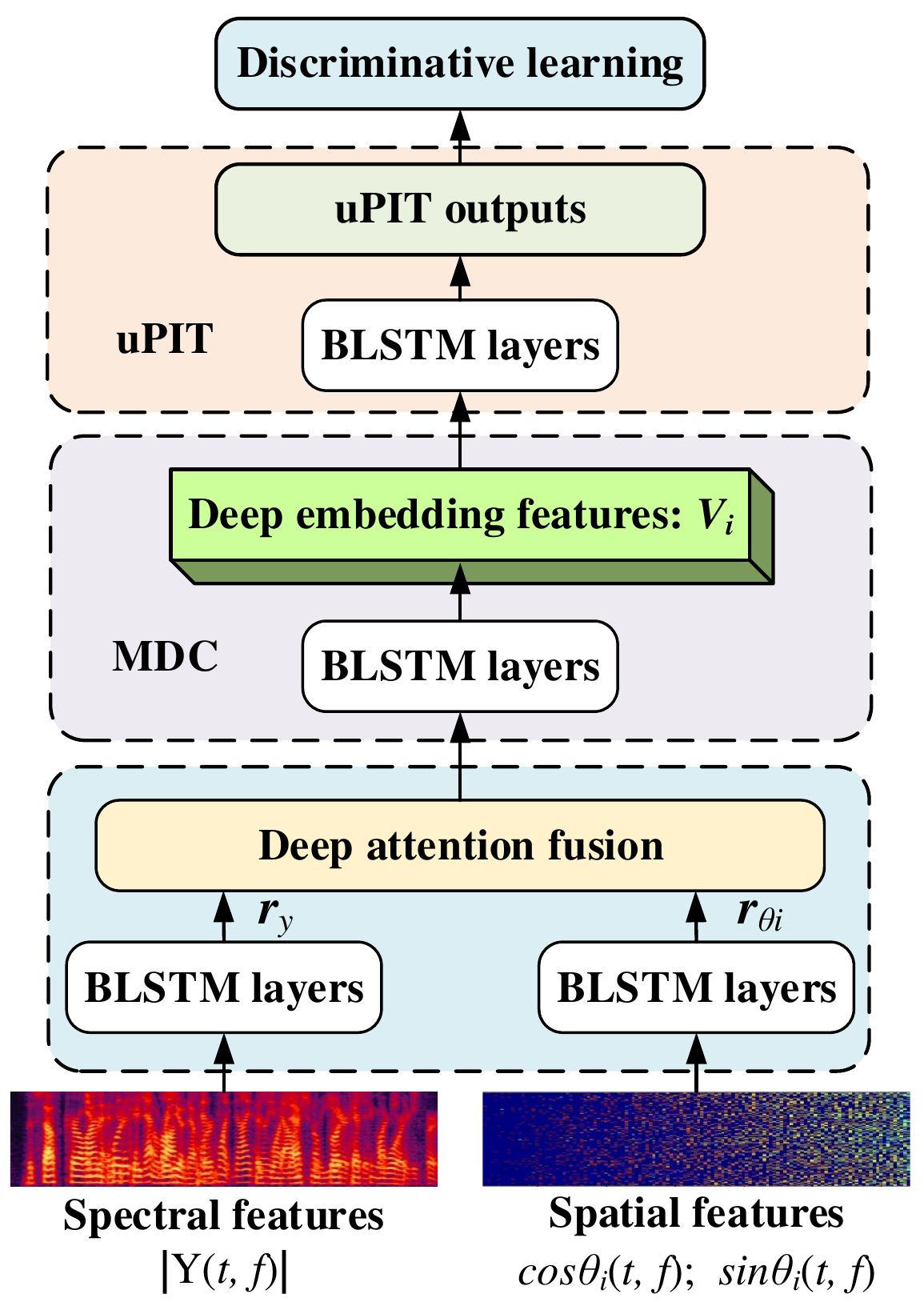}
	\end{minipage}
	\caption{Schematic diagram of our proposed multi-channel speech separation method.}
	\label{fig:MDC_DEF}
\end{figure}


In this section, we present our proposed spatial and spectral deep attention fusion algorithm for speech separation using deep embedding features, which is shown in Fig.~\ref{fig:MDC_DEF}. Instead of simply stacking the spatial and spectral features, the deep attention fusion algorithm is utilized to combine them deeply, which uses an attention model to dynamically control the weights of the spectral and spatial features. Therefore, the separation model can learn the mutual relationship between the spectral and spatial features. In addition, to address the training objective problem of MDC, the real separated sources are used as the training objectives, which uses the deep embedding features for multi-channel speech separation.

\vspace{-1ex}
\subsection{Deep Attention Fusion}

As shown in Fig.~\ref{fig:MDC_DEF}, the spectral features \(|Y(t,f)|\) and spatial IPDs \(\{cos\theta_i(t,f); sin\theta_i(t,f)\}\) are firstly processed by the BLSTM network to acquire the deep representations. These spectral and spatial deep representations are denoted by \(\textbf{\textsl{r}}_y\) and \(\textbf{\textsl{r}}_{\theta_i}\), respectively. In order to make the separation model dynamically control the weights of the spectral and spatial features, an attention module \cite{bahdanau2014neural} is applied between the \(\textbf{\textsl{r}}_y\) and \(\textbf{\textsl{r}}_{\theta_i}\). 

According to the attention mechanism \cite{bahdanau2014neural}, the attention weight \(\alpha_{t,t^{'}}\) can be learned:
\begin{equation}
\alpha_{t,t^{'}}=\frac{{\rm exp}(d_{t,t^{'}})}{\sum_t^{'}{{\rm exp}(d_{t,t^{'}})}}
\label{eq3}
\end{equation}
where \(d_{t,t^{'}}\) is the mutual relationship between \(\textbf{\textsl{r}}_y\) and \(\textbf{\textsl{r}}_{\theta_i}\), which measures their importances for speech separation. The attention weight \(\alpha_{t,t^{'}}\) is the softmax of \(d_{t,t^{'}}\) over \(t^{'}\in{[1,T]}\) (\(T\) denotes the number of frames). \(d_{t,t^{'}}\) is defined as follows:
\begin{equation}
d_{t,t^{'}}=\textbf{\textsl{r}}_y^{T}\textbf{\textsl{r}}_{\theta_i}
\label{eq4}
\end{equation}

The context vector \(\textbf{c}_{t}\) can be calculated by the weighted average of \(\textbf{\textsl{r}}_{\theta_i}\):
\begin{equation}
\textbf{c}_{t}=\sum_{t^{'}}{\alpha_{t,t^{'}}\textbf{\textsl{r}}_{\theta_i}}
\label{eq5}
\end{equation}
The context vector \(\textbf{c}_{t}\) and these spectral and spatial deep representations (\(\textbf{\textsl{r}}_y\) and \(\textbf{\textsl{r}}_{\theta_i}\)) are used as the deep attention fusion features. They are applied to extract the deep embedding features.

\vspace{-2ex}
\subsection{Deep Embedding Features for Separation}

Clusters in the embedding space of MDC can represent the inferred spectral masking patterns of individual sources. In this paper, we utilize the MDC network to extract deep embedding features, which contain the information of each source and are conducive to speech separation. The D-dimensional deep embedding features \(V_i\) (\(i\) is the index of a microphone pair) can be extracted as follows:
\begin{equation}
V_i=\xi_{BLSTM}{\{\textbf{\textsl{r}}_y; \textbf{c}_{t}; \textbf{\textsl{r}}_{\theta_i}\}} \ (i=1,2,...,N_{m}-1)
\label{eq6}
\end{equation}
where \(\xi_{BLSTM}\) denotes the mapping of BLSTM network.

In order to address the training objective problem of MDC, instead of using the unsupervised K-means clustering, the supervised uPIT network is applied to estimate soft masks from these deep embedding features. When the \(V_1, V_2, ..., V_{N_m-1}\) are stacked, they are sent to the uPIT network: \(\zeta_{uPIT}=\{V_1, V_2, ..., V_{N_m-1}\}\). The uPIT network computes the mean square error (MSE) for all possible speaker permutations at utterance-level. Then the minimum cost among all permutations (P) is chosen as the optimal assignment.
\begin{equation}
\phi^*=\mathop{arg\,min}_{{P}}\sum_{s=1}^S|||Y|\odot{\widetilde{M}_s}-|X_{s}|cos(\theta_y-\theta_s)||_F^2
\label{eq7}
\end{equation}
where the number of all permutations P is \(S!\) (\(!\) denotes the factorial symbol). \(\widetilde{M}_s\) is the estimated phase sensitive mask (PSM) \cite{wang2018supervised} of source \(s\). \(\theta_y\) and \(\theta_s\) are the reference microphone phase of mixture speech and target source \(s\). \(|X_{s}|\) is the spectrogram of target source \(s\).

\vspace{-1ex}
\subsection{Discriminative Learning and Joint Training}

To reduce the distance between the same speakers and increase the distance between different speakers, discriminative learning (DL) is applied to our proposed model. The DL loss function can be defined as follows:
\begin{equation}
J_{DL}=\phi^*-\sum_{\phi \ne{\phi^*},\phi \in{P}}\alpha \phi
\label{eq8}
\end{equation}
where \(\phi\) is a permutation from \(P\) but not \(\phi^*\), \(\alpha \ge0\) is the regularization parameter of \(\phi\). When \(\alpha=0\), the loss function is same as the \(\phi^*\) in Eq.~\ref{eq7}. It means without DL.

To extract embedding features effectively, we apply the joint training framework to the proposed system. The loss function of joint training is defined as follows:
\begin{equation}
J =\lambda{J_{DC}}+(1-\lambda)J_{DL}
\label{eq9}
\end{equation}
where \(\lambda\in{[0,1]}\) controls the weight of \(J_{DC}\) and \(J_{DL}\).

\vspace{-1ex}
\section{Experiments and Results}

\vspace{-1ex}
\subsection{Dataset}


The room impulse response (RIR) generator \footnote{Available online at https://github.com/ehabets/RIR-Generator} is used to spatialize the WSJ0-2mix dataset \cite{Hershey2016Deep}. The dataset consists of three sets: training set (20,000 utterances about 30 hours), validation set (5,000 utterances about 10 hours) and test set (3,000 utterances about 5 hours). Specifically, the training and validation sets are generated by randomly selecting utterances from WSJ0 training set (\texttt{si\_tr\_s}). Similar as generating training and validation set, the test set is created by mixing the utterances from the WSJ0 development set (\texttt{si\_dt\_05}) and evaluation set (\texttt{si\_et\_05}).

The RIR is generated using the image-source method \cite{allen1979image} with a linear microphone array with 4 microphones. The reverberation time \(RT_{60}\) is set to 0.16s. The distances between 4 microphones are 4-8-4 cm. For any two speakers, we constrain them to be at least \(45\degree\) apart. We mix the images of two speakers with signal-to-noise ratios (SNRs) between -5dB and 5dB. The average distance between a source and array center is set to 1m.

\setlength{\tabcolsep}{0.6mm}{
	\begin{table*}[t]
		\caption{The results of SDR, PESQ and STOI for different separation methods and different gender combinations. The ``{MDC+attention}" means that the MDC applies the deep attention fusion algorithm for spatial and spectral features. The ``{Proposed(IAM)}" and ``{Proposed(IPSM)}" mean that the proposed method uses the IAM and the IPSM as the output, respectively.}
		\label{tab:results1}
		\begin{tabular}{c|ccc|ccc|ccc|ccc}
			\hline
			\multirow{2}{*}{Methods} & \multicolumn{3}{c|}{Male-Female}               & \multicolumn{3}{c|}{Female-Female}             & \multicolumn{3}{c|}{Male-Male}                 & \multicolumn{3}{c}{AVG.}                      \\ \cline{2-13} 
			& SDR(dB)       & PESQ          & STOI(\%)       & SDR(dB)       & PESQ          & STOI(\%)       & SDR(dB)       & PESQ          & STOI(\%)       & SDR(dB)       & PESQ          & STOI(\%)       \\ \hline
			Mixture                  & 0.15          & 1.32          & 61.26          & 0.16          & 1.37          & 62.38          & 0.15          & 1.30          & 61.74          & 0.15          & 1.33          & 61.59          \\ \hline
			MDC(baseline)            & 12.7          & 2.70          & 89.75          & 13.0          & 2.81          & 90.95          & 11.9          & 2.55          & 89.67          & 12.5          & 2.68          & 89.94          \\ \hline
			MDC+attention            & 12.9          & 2.72          & 89.88          & 13.3          & 2.83          & 91.08          & 12.0          & 2.58          & 89.83          & 12.7          & 2.70          & 90.01          \\ 
			Proposed(IAM)        & 12.9          & \textbf{3.58} & \textbf{94.17} & 13.3          & \textbf{3.60} & \textbf{95.32} & 11.9          & \textbf{3.55} & \textbf{94.34} & 12.7          & \textbf{3.58} & \textbf{94.42} \\ 
			Proposed(IPSM)       & \textbf{14.5} & 3.39          & 93.52          & \textbf{15.2} & 3.44          & 95.23          & \textbf{13.8} & 3.32          & 93.64          & \textbf{14.5} & 3.38          & 93.86          \\ \hline
			IBM                      & 13.7          & 3.29          & 91.77          & 14.2          & 3.36          & 93.75          & 12.8          & 3.17          & 91.05          & 13.5          & 3.26          & 91.91          \\ 
			IAM                      & 13.0          & 3.80          & 95.44          & 13.4          & 3.79          & 96.38          & 12.0          & 3.82          & 95.19          & 12.8          & 3.80          & 95.53          \\ 
			IPSM                     & {16.7} & {4.05} & {96.40} & {17.1} & {4.05} & {97.15} & {15.7} & {4.04} & {95.71} & {16.5} & {4.05} & {96.33} \\ \hline
		\end{tabular}
	\end{table*}
}

\vspace{-2ex}
\subsection{Experimental setup}
\vspace{-1ex}


The first channel is used as the reference microphone. The sampling rate of all generated data is 8 kHz. The short-time Fourier transform (STFT) has 32 ms length hamming window and 8 ms window shift.

The proposed method contains 4 BLSTM layers, each with 600 units in each direction. More specifically, as for the deep attention fusion module, there is only one BLSTM layer for spectral and spatial, respectively. As for the DC network, there is also only one BLSTM layer. As for the uPIT network, there are two BLSTM layers. The dimension D of the embeddings is set to 20 per T-F bin \cite{wang2018multi}. The regularization parameter \(\alpha\) of discriminative learning is set to 0.1. And the joint training weight \(\lambda\) is set to 0.01.

We apply the MDC \cite{wang2018multi} as our baseline and re-implement it with our experimental setup. MDC has 4 BLSTM layers with 600 units, which is same as the proposed method.



In this work, the models are evaluated on the signal-to-distortion ratio (SDR) \cite{vincent2006performance}, the perceptual evaluation of speech quality (PESQ) \cite{rix2002perceptual} measure and the short-time objective intelligibility (STOI) measure \cite{taal2010short}.


\vspace{-2ex}
\subsection{Experimental results}
\vspace{-1ex}

Table~\ref{tab:results1} shows the results of SDR, PESQ and STOI for different separation methods and different gender combinations. The ``\textbf{MDC+attention}" means that the MDC applies the deep attention fusion algorithm for spatial and spectral features. The ``\textbf{Proposed(IAM)}" and ``\textbf{Proposed(IPSM)}" mean that the proposed method uses the ideal amplitude mask (IAM) \cite{Kolbaek2017Multitalker} and the ideal PSM (IPSM) as the output, respectively. In addition, the last three rows present the results of the ideal binary mask (IBM), IAM and IPSM, which are oracle masks.

\vspace{-2ex}
\subsubsection{Evaluation of deep attention fusion}

From Table~\ref{tab:results1} we can find that when the deep attention fusion is applied to the MDC, the performance of speech separation can be improved no matter what gender combinations for these three evaluation metrics. This result indicates that this deep attention fusion algorithm is effective for speech separation. The reason is that this deep attention fusion algorithm utilizes an attention module to dynamically control the weights of the spectral and spatial features. Therefore, this deep attention fusion algorithm can make the separation model pay more attention to spectral or spatial according to their contributions. In other words, it combines the spectral and spatial features deeply. So compared with the MDC method, the proposed ``MDC+attention" method can acquire a better speech separation performance.

\vspace{-2ex}
\subsubsection{The effectiveness of our proposed method}
\vspace{-1ex}

From Table~\ref{tab:results1} we can make several observations. Firstly, compared with the MDC baseline method, the performance of the proposed methods can be largely improved. More specifically, compared with the MDC, the proposed IPSM based method obtains 16.0\%, 26.1\% and 4.4\% relative improvements in SDR, PESQ, and STOI, respectively. Secondly, we surprisingly find that the results of the proposed IPSM based method are better than using the oracle mask of IBM. Note that these IBM results are the limit results of MDC baseline method. These results indicate the effectiveness of the proposed method. Thirdly, although the SDR results of the proposed IAM method are slightly better than the baseline, the PESQ and STOI results are significantly better than the baseline. These is because that the SDR results of oracle IAM are worse than the IBM and IPSM but the PESQ and STOI results of oracle IAM are better than IBM and comparable to IPSM. Finally, no matter what gender combinations, our proposed speech separation method can acquire better results than the baseline method. These results reveal that our proposed method has an effective ability to reconstruct target sources for all of the gender combinations.

\begin{figure}[t]
	\centering
	\begin{minipage}[t]{0.45\textwidth}
		\includegraphics[width=\linewidth]{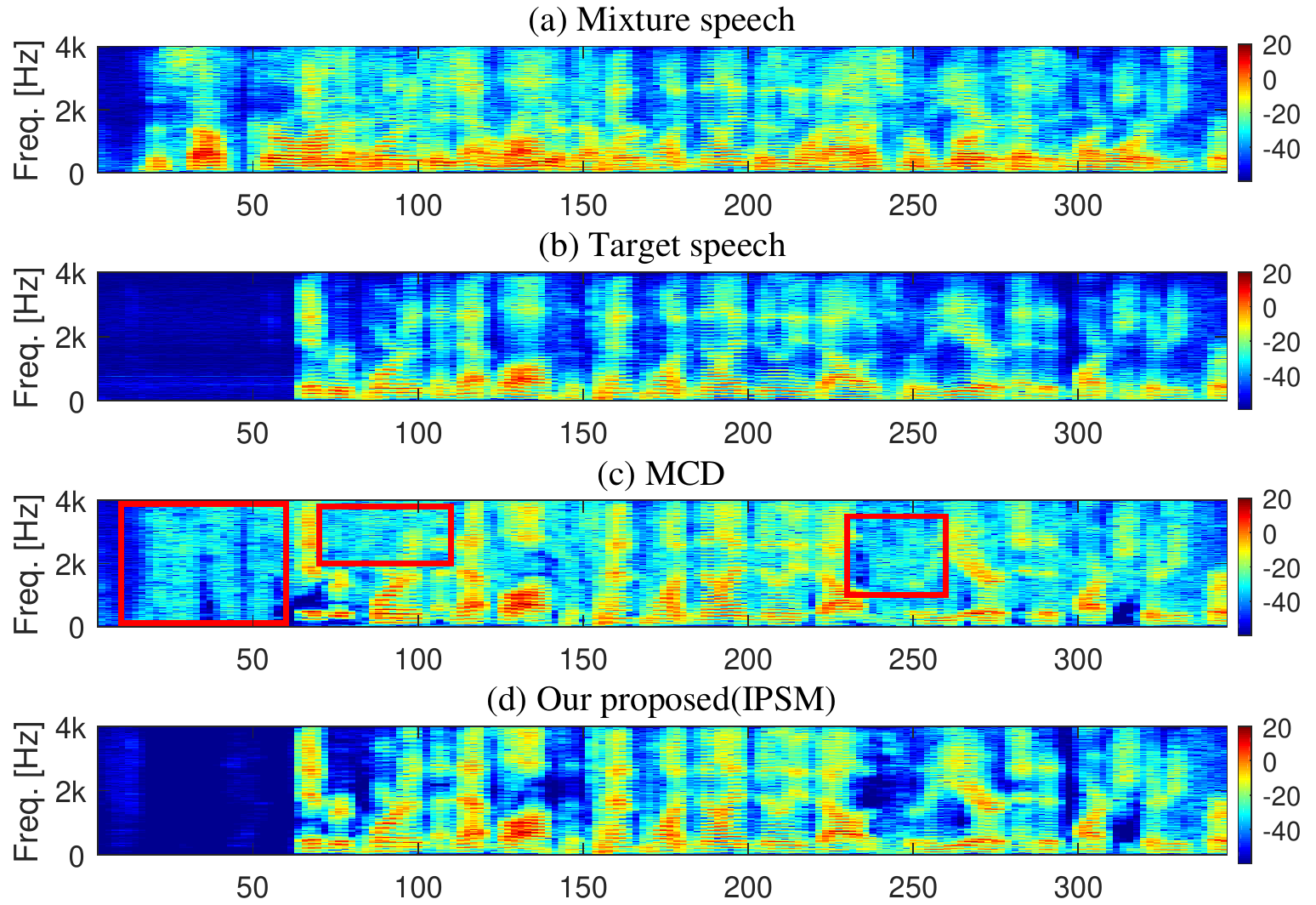}
	\end{minipage}
	\caption{The spectrogram example of different speech separation methods. (a) The mixture speech. (b) The target speech. (c) The separated speech of MDC. (d) The separated speech of our proposed(IPSM) method.}
	\label{fig:yuputu}
\end{figure}

Fig.~\ref{fig:yuputu} shows a spectrogram example of different separation methods. Notice that, although the baseline MDC can separate the mixture, some interference signals are not removed very well (marked in the red boxes). Compared with the spectrogram of target speech, the harmonics and formant structures of the proposed model are effectively preserved in the reconstructed speech. In addition, the interference signals are removed very well. These results indicate that the mixture speech is effectively separated by the proposed model.

\vspace{-2ex}
\section{Conclusions}
\vspace{-1ex}

In this paper, we propose a spatial and spectral deep attention fusion method for multi-channel speech separation using deep embedding features. In order to dynamically control the weight of the spectral and spatial features, the attention module is applied to compute the importance of the spectral and spatial features for speech separation. In addition, to address the training objective problem of MDC, the MDC is applied to extract deep embedding features. Instead of utilizing the unsupervised K-means clustering, the supervised uPIT network is used to learn soft target masks. Results show that the proposed method outperforms MDC baseline, with relative improvements of 16.0\%, 26.1\% and 4.4\% in SDR, PESQ, and STOI, respectively. Besides, the proposed method is even better than the oracle IBM. 


%

\vfill\pagebreak

\bibliographystyle{IEEEbib}
\bibliography{refs}

\end{document}